\documentstyle{mn}
\font\japit = cmti10 at 10truept
\input{epsf.sty}

\title
     [Optimal Power Spectra I]
{\vglue-3.0truecm
\centerline{\japit Submitted to Monthly Notices}
\vglue 2.5truecm
\noindent
Towards Optimal Measurement of Power Spectra I: \\
Minimum Variance Pair Weighting and the Fisher Matrix
\author
     [A. J. S. Hamilton]
     {A. J. S. Hamilton \\
	JILA and Dept.\ of Astrophysical, Planetary and Atmospheric Sciences,
	Box 440, University of Colorado, Boulder CO 80309, USA\\
	email: ajsh@dark.colorado.edu}}

\newcommand{\rmn}{\rm}
\newcommand{\bmi}{\bmath}
\newcommand{\be}{\begin{equation}}
\newcommand{\ee}{\end{equation}}
\newcommand{\ba}{\begin{eqnarray}}
\newcommand{\ea}{\end{eqnarray}}
\newcommand{\nn}{\nonumber \\}
\newcommand{\dd}{{\rmn d}}	% MNRAS
\newcommand{\e}{{\rmn e}}	% MNRAS
\newcommand{\im}{{\rmn i}}	% MNRAS
\newcommand{\ddd}{\dd^3\!}
\newcommand{\up}[2]{\raisebox{0pt}[2.4ex]{\mathsurround=0pt
		\begin{array}[b]{@{}l@{}} \; \scriptstyle{#1} \\[-1ex]
		#2 \end{array}}}
\newcommand{\Sym}[1]{\begin{array}[t]{@{\;}c@{\:}} {\rmn Sym}
		\\[-1ex] \scriptstyle #1 \end{array}}

\newcommand{\k}{{\bmi k}}

\newcommand{\n}{{\bmi n}}
\newcommand{\r}{{\bmi r}}

\newcommand{\ff}{\beta}
\newcommand{\G}{G}
\newcommand{\PI}{\pi}

\newcommand{\XI}{C}

\begin{document}

\maketitle

\begin{abstract}
This is the first of a pair of papers which address the
problem of measuring the unredshifted power spectrum
in optimal fashion from a survey of galaxies,
with arbitrary geometry,
for Gaussian or non-Gaussian fluctuations,
in real or redshift space.
In this first paper,
that pair weighting is derived which formally minimizes the expected variance
of the unredshifted power spectrum windowed over some arbitrary kernel.
The inverse of the covariance matrix
of minimum variance estimators of windowed power spectra
is the Fisher information matrix,
which plays a central role in establishing optimal estimators.
Actually computing the minimum variance pair window and the Fisher
matrix in a real survey still presents a formidable numerical problem,
so here a perturbation series solution is developed.
The properties of the Fisher matrix evaluated according to the approximate
method suggested here are investigated in more detail in the second paper.
\end{abstract}


%\clearpage

\section{Introduction}
\label{intro}

The problem of measuring the power spectrum of fluctuations in the Universe
in optimal fashion has received some attention recently
(Vogeley \& Szalay 1996;
Tegmark, Taylor \& Heavens 1997;
and references therein).
Vogeley \& Szalay
argue that the Karhunen-Lo\`eve (KL) basis,
the basis of signal-to-noise eigenmodes of the survey covariance,
provides the optimal basis for measuring the power spectrum.
Tegmark et al.\ give an elegant review of what constitutes
optimal measurement of parameters,
and discuss how to compress (reweight and truncate) a KL basis
to tractable size when dealing with large data sets
such as the forthcoming 2df survey,
or the Sloan Digital Sky Survey.

Tegmark et al.\ emphasize the central role of the Fisher information matrix,
which is the matrix of second derivatives of minus the log-likelihood function
with respect to the parameters,
and whose inverse gives an estimate of the covariance matrix of the parameters.
Tegmark et al.\ conclude, in the final sentence of their \S4, that
``In general, it does not seem tractable to devise a data-weighting
scheme which optimizes the Fisher matrix directly''.
A principal aim of the present paper is to demonstrate that
the problem is after all tractable, albeit in an approximation,
for the case where the parameter set to be measured is the unredshifted
power spectrum of galaxies.
The paper which follows this one (Hamilton 1997, hereafter Paper~2)
investigates in more detail the properties of the approximate Fisher matrix
computed according to the procedure described in the present paper.

I suppose that one has at hand a galaxy survey, in real or redshift space,
in which the observed galaxy density at position $\r$ is $N(\r)$,
and the selection function, the expected density of galaxies at $\r$,
is a function $\Phi(\r)$ which is known or can be measured with
negligible uncertainty.
The observed overdensity $\delta(\r)$ at $\r$ is then defined to be
\be
\label{delta}
  \delta(\r) \equiv {N(\r) - \Phi(\r) \over \Phi(\r)}
  \ .
\ee
The overdensity $\delta$ here is taken to be defined in real (unredshifted)
space,
but it is shown in \S\ref{redshift} how the results generalize
to the case of linear distortions in redshift space.

The basic problem considered in this paper is to estimate
the true unredshifted galaxy covariance function
$\xi = \langle \delta \delta \rangle$
from the data in real or redshift space.
The true covariance $\xi$ of galaxies in the Universe
differs from the observed covariance
because of noise and because of finiteness of the sample.
In the real representation the true unredshifted covariance function
is the correlation function $\xi(r)$,
which on the assumption of statistical homogeneity and isotropy is
a function of scalar separation $r$,
while in the Fourier representation it is the power spectrum $\xi(k)$,
a function of scalar wavenumber $k$
%(I adopt the symmetric normalization of Fourier transforms in this paper,
%to avoid extraneous factors of $2\PI$)
\be
\label{xik}
% Symmetric:
%  \xi(k) = (2\PI)^{-3/2} \int \e^{\im \k.\r} \xi(r) \, \ddd r
%    = (2/\PI)^{1/2} \int_0^\infty \! j_0(kr) \xi(r) r^2 \dd r
% Asymmetric:
  \xi(k) = \int \e^{\im \k.\r} \xi(r) \, \ddd r
    = \int_0^\infty \! j_0(kr) \xi(r) \, 4\PI r^2 \dd r
\ee
where $j_0(x) = \sin(x)/x$ is a spherical Bessel function.
I adhere throughout this paper to what has become the standard convention
in cosmology for defining Fourier transforms,
notwithstanding the extraneous factors of $2\PI$ which result.

I start by considering, in \S\ref{minvariance},
the power spectrum
$\xi(k)$ windowed over some arbitrary prescribed kernel $\G(k)$,
\be
\label{Gxik}
% Symmetric:
%  \tilde\xi \equiv \int \G(k) \xi(k) \, 4\PI k^2 \dd k
% Asymmetric:
  \tilde\xi \equiv \int \G(k) \xi(k) \, 4\PI k^2 \dd k /(2\PI)^3
  \ .
\ee
The windowed power spectrum $\tilde\xi$
can be estimated by an estimator $\hat\xi$
(the hat distinguishing the estimator from the true value $\tilde\xi$)
which is quadratic in observed overdensities $\delta$:
\be
\label{xiest}
  \hat\xi
    = \int W(\r_i,\r_j) \delta(\r_i) \delta(\r_j) \, \ddd r_i \ddd r_j
  \ .
\ee
Section~\ref{minvariance}
derives an expression,
equation~(\ref{minvar}),
for that pair weighting $W^{ij}$ of overdensities $\delta_i \delta_j$
which formally minimizes the expected variance
amongst quadratic estimators $\hat\xi$
of the windowed power spectrum $\tilde\xi$.

The inverse $T^{\alpha \beta}$ of the expected covariance
$\langle \Delta\hat\xi_\alpha \Delta\hat\xi_\beta \rangle$
between minimum variance estimators of the power spectrum
is the Fisher information matrix for the power spectrum,
as discussed in \S\ref{fisher}.
In effect,
the Fisher matrix defines the maximum amount of information
that can be extracted about the power spectrum from a given survey,
and is therefore a fundamental quantity in designing optimal procedures
for measuring the power spectrum.
Paper~2 shows how the Fisher matrix can be used to construct
a complete set of positive kernels yielding
a complete statistically orthogonal set of windowed power spectra.

Actually evaluating the Fisher matrix $T^{\alpha \beta}$
still presents a formidable problem,
involving inversion of the covariance
$\langle \Delta\XI_{ij} \Delta\XI_{kl} \rangle$
of the survey covariance
(eq.~[\ref{XiXiNP}] below),
which is a rank 4 matrix of 3-dimensional quantities.
A solution to the problem is proposed
in \S\S\ref{classical} and \ref{perturb}.
In \S\ref{classical},
it is shown that for Gaussian fluctuations,
in the classical limit where the wavelength is short
compared to the scale of the survey,
the minimum variance pair window goes over
to that derived by Feldman, Kaiser \& Peacock (1994, hereafter FKP).
The problem of generalizing the FKP pair window
to the non-Gaussian case is addressed in \S\ref{classicalng},
but unfortunately
I have been unable to find an elegant way to implement such a generalization.
In \S\ref{perturb},
a perturbation series solution of the Fisher matrix is developed,
starting from the classical FKP solution in the Gaussian case.
A perturbation solution exists also in the non-Gaussian case,
\S\ref{perturbng},
but the lack of a simple non-Gaussian generalization of the
classical FKP window makes the choice of zeroth order solution here less clear.
Paper~2 evaluates the Fisher matrix assuming Gaussian fluctuations
and the zeroth order FKP approximation.

Section~\ref{approximate} offers advice on implementing
an approximate minimum variance pair weighting.
Section~\ref{summary} summarizes the conclusions.

\section{Minimum Variance Weighting}
\label{minvariance}

\subsection{Prior}
\label{prior}

To derive minimum variance measures,
it is necessary to make prior assumptions about the origin of the
variance in a survey.
Such prior assumptions can be regarded as priors in Bayesian model testing,
or alternatively as reasonable guesses which in practice may yield
near minimum variance estimates.

In the first place,
I assume that the expectation value of the
(unredshifted) survey covariance
is the sum of the cosmic covariance $\xi$ and a Poisson sampling term
\be
\label{Xi}
  \XI(\r_i,\r_j)
    \equiv \langle \delta(\r_i) \delta(\r_j) \rangle
    = \xi(r_{ij}) + \delta_D(\r_{ij}) \Phi(\r_i)^{-1}
\ee
where $r_{ij} \equiv | \r_{ij} |$ and $\r_{ij} \equiv \r_i-\r_j$,
and $\delta_D$ denotes a Dirac-delta function.
This assumption implies in particular that the (unredshifted) survey covariance
$\XI(\r_i,\r_j)$ at any pair of non-coincident points
$\r_i \neq \r_j$
provides an estimate of the cosmic covariance $\xi(r_{ij})$
\be
\label{XiNP}
  \XI(\r_i,\r_j)_{\r_i \neq \r_j}
    = \xi(r_{ij})
  \ .
\ee
Secondly, I suppose that the expected covariance
$\langle \Delta\XI_{ij} \Delta\XI_{kl} \rangle$
of the survey covariance function $\XI_{ij}$ is a specified function.
Under the same assumption that it
is a sum of cosmic and Poisson sampling terms,
the covariance
$\langle \Delta\XI_{ij} \Delta\XI_{kl} \rangle$
takes the general form
\be
\label{XiXi}
  \langle \Delta\XI_{ij} \Delta\XI_{kl} \rangle
    = \XI_{ik} \XI_{jl} + \XI_{il} \XI_{jk} + {\cal H}_{ijkl}
\ee
where ${\cal H}_{ijkl}$ is the expected 4-point correlation function of the
survey, which is a sum of the cosmic 4-point function $\eta_{ijkl}$
plus Poisson sampling terms where any two or more of the points $ijkl$ coincide,
\ba
  {\cal H}_{ijkl}
  \!\!\! &=& \!\!\!
    \eta_{ijkl}
    + \bigl[ \delta_D(\r_{ij}) \zeta_{ikl} \Phi_i^{-1} + \mbox{cyc.} \bigr]
    (\mbox{6 terms})
  \nn
  && \!\!\!
  \mbox{}
    + \bigl[ \delta_D(\r_{ij}) \delta_D(\r_{kl})
    \xi_{ik} \Phi_i^{-1} \Phi_k^{-1}
    + \mbox{cyc.} \bigr] (\mbox{3 terms})
  \nn
  && \!\!\!
  \mbox{}
    + \bigl[ \delta_D(\r_{ij}) \delta_D(\r_{ik}) \xi_{il} \Phi_i^{-1}
    + \mbox{cyc.} \bigr] (\mbox{4 terms})
  \nn
  && \!\!\!
  \mbox{}
    + \delta_D(\r_{ij}) \delta_D(\r_{ik}) \delta_D(\r_{il}) \Phi_i^{-1}
    (\mbox{1 term})
\label{HH}
\ea
with $\zeta_{ijk}$ the 3-point function.
In estimating the cosmic covariance $\xi(r_{ij})$ from
the survey covariance $\XI(\r_i,\r_j)$,
it is necessary to omit (or subtract off) the Poisson noise term where $i = j$,
as in equation~(\ref{XiNP}).
The covariance of the survey covariance with these Poisson terms removed
is given by
\be
\label{XiXiNP}
  \langle \Delta\XI_{ij} \Delta\XI_{kl} \rangle_
    {\begin{array}[c]{@{}l@{}}
      \scriptstyle{i\neq j} \\[-1ex] \scriptstyle{k \neq l}
    \end{array}}
    = \XI_{ik} \XI_{jl} + \XI_{il} \XI_{jk} + H_{ijkl}
\ee
where the 4-point function $H_{ijkl}$
includes only those Poission sampling terms where $i$ or $j$ = $k$ or $l$,
not those where $i = j$ or $k = l$,
\ba
  H_{ijkl}
  \!\!\! &=& \!\!\!
    \eta_{ijkl}
    + \bigl[ \delta_D(\r_{ik}) \zeta_{ijl} \Phi_i^{-1}
    + (i \leftrightarrow j, k \leftrightarrow l) \bigr] (\mbox{4 terms})
  \nn
  && \!\!\!\!\!\!
  \mbox{}
    + \bigl[ \delta_D(\r_{ik}) \delta_D(\r_{jl})
    \xi_{ij} \Phi_i^{-1} \Phi_j^{-1}
    + (k \leftrightarrow l ) \bigr] (\mbox{2 terms})
  \ .
  \nn
\label{H}
\ea

In the particular case of Gaussian fluctuations,
the 4-point function ${\cal H}_{ijkl}$ (and likewise $H_{ijkl}$) vanishes,
in which case the covariance of the survey covariance reduces to
\be
\label{XiXig}
  \langle \Delta\XI_{ij} \Delta\XI_{kl} \rangle
    = \XI_{ik} \XI_{jl} + \XI_{il} \XI_{jk}
  \ .
\ee
Although the results of this paper are not confined to the Gaussian case,
the combination of
poor knowledge of the 4-point function,
the additional complication of the non-Gaussian case,
and the expectation that fluctuations may well be Gaussian on linear scales,
means that the Gaussian hypothesis is a natural first choice for model testing,
at least on sufficiently large scales.

\subsection{Notation}
\label{notation}

Henceforward,
it is convenient to adopt an abbreviated notation in which
Latin indices $ij$ denote pairs of volume elements
at positions $\r_i$ and $\r_j$ in a survey,
while Greek indices $\alpha$ denote pair separations $r_\alpha$.
Introduce the convention that replacing a pair index $ij$
on a vector or matrix by a separation index $\alpha$
means sum over all pairs $ij$ separated by $r_\alpha$ in the survey.
Thus, in the real space representation,
\be
\label{reduce}
  W^\alpha \equiv \int W^{ij} \delta_D (|\r_i-\r_j|-r_\alpha)
    \, \ddd r_i \ddd r_j
\ee
and similarly for matrices with more indices.
Normalizing the Dirac delta-function
in equation~(\ref{reduce}) on a 3-dimensional scale so that
$\int \delta_D (r_{ij}-r_\alpha) \discretionary{}{}{}
4\PI r_\alpha^2 \dd r_\alpha =
\int \delta_D (r_{ij}-r_\alpha) \discretionary{}{}{}
4\PI r_{ij}^2 \dd r_{ij} = 1$
ensures the usual interpretation of, for example,
$\G^\alpha \xi_\alpha = \int \G(r) \xi(r) 4\PI r^2 \dd r$
(see equation~[\ref{Gxi}])
in the real space representation.

To exhibit the derivation of the minimum variance pair weighting
in as general a form as possible,
it is useful to borrow ideas from quantum mechanics,
and to treat all quantities,
such as the covariance function $\xi_\alpha$,
or the pair weighting $W^{ij}$, as vectors in a Hilbert space.
Such vectors have a meaning independent of the particular basis
with respect to which they might be expressed.
For example, the covariance function is the correlation function $\xi(r)$
in the real space representation,
or the power spectrum $\xi(k)$ in the Fourier representation,
but from a Hilbert space point of view these quantities are the same vector,
and in this paper I use the same symbol $\xi$ to denote them both.
(It would be nice to adopt the Dirac bra-ket notation,
but unfortunately this leads to confusion with ensemble averages.)

To be explicit,
suppose that $\psi_\alpha(r)$ constitute
some arbitrary complete set of orthonormal separation functions,
labelled by separation index $\alpha$.
Then the covariance function $\xi_\alpha$ in the $\psi$-representation
is related to $\xi(r)$ in the real space representation by
\be
\label{xiar}
  \xi_\alpha = \int \psi_\alpha(r) \xi(r) \, 4\PI r^2 \dd r
  \ ,
\ee
\be
\label{xira}
  \xi(r) = \xi_{\alpha} \psi^{\alpha}(r)
  \ .
\ee
The summation convention for repeated indices is adopted in~(\ref{xira})
and hereafter.
The raised index on $\psi^\alpha$ denotes the Hermitian conjugate of
$\psi_\alpha$;
one of a pair of repeated indices is always raised,
the other lowered.
The orthonormality conditions on $\psi_\alpha$ are
\be
\label{psiortha}
  \int \psi_\alpha(r) \psi_\beta(r) \, 4\PI r^2 \dd r
    = \delta_{D \alpha \beta}
\ee
where
$\delta_{D \alpha \beta}$ denotes the unit matrix in $\psi$-space
(the subscript $D$ is retained to distinguish the
unit matrix $\delta_D$ from the overdensity $\delta$;
for discrete representations,
$\delta_D$ should be interpreted as a Kronecker delta rather than
a Dirac delta-function);
and
\be
\label{psiorthr}
  \psi^\alpha(r_1) \psi_\alpha(\r_2)
    = \delta_D(\r_{12})
  \ .
\ee
Similarly,
suppose that $\phi_{ij}(\r_1,\r_2)$ constitute
some arbitrary complete orthonormal set of pair functions,
which by pair exchange symmetry may be taken to be symmetric in
$i \leftrightarrow j$ and $\r_1 \leftrightarrow \r_2$
without loss of generality.
Then a pair function $W^{ij}$ in the $\phi$-representation is
related to $W(\r_1,\r_2)$ in the real space representation by
\be
  W^{ij} = \int \phi^{ij}(\r_1,\r_2) W(\r_1,\r_2) \, \ddd r_1 \ddd r_2
\ee
\be
  W(\r_1,\r_2) = W^{ij} \phi_{ij}(\r_1,\r_2)
  \ .
\ee
Define the symbol Sym to symmetrize over its underscripts, as in
\be
\label{sym}
  \! \Sym{(ij)} C_{ij} \equiv (C_{ij} + C_{ji})/2
  \ .
\ee
The orthonormality conditions on the pair functions $\phi_{ij}$ are
\be
  \int \phi_{ij}(\r_1,\r_2) \phi_{kl}(\r_1,\r_2) \, \ddd r_1 \ddd r_2
    = \Sym{(kl)} \delta_{Dik} \delta_{Djl}
\ee
where
${\rmn Sym}_{(kl)} \delta_{Dik} \delta_{Djl}$
is the unit matrix in $\phi$-space,
and
\be
  \phi^{ij}(\r_1,\r_2) \phi_{ij}(\r_3,\r_4)
    = \Sym{(34)} \delta_D(\r_{13}) \delta_D(\r_{24})
\ee
where again
${\rmn Sym}_{(34)} \delta_D(\r_{13}) \delta_D(\r_{24})$
is the unit matrix for pairs in the real representation.

In the real space representation,
the basis of orthonormal separation functions is the set of
delta-functions in real space,
$\psi_\alpha(r) = \delta_D(r{-}r_\alpha)$,
and summation over $\alpha$ signifies integration over
$4 \PI r_\alpha^2 \dd r_\alpha$.
The basis of pair functions in the real representation
is similarly the set of pairs of delta-functions in real space,
$\phi_{ij}(\r_1,\r_2)
= {\rmn Sym}_{(ij)} \delta_D(\r_1{-}\r_i) \delta_D(\r_2{-}\r_j)$,
and summation over $ij$ signifies integration over $\ddd r_i \ddd r_j$.
In the real space representation,
the Hermitian conjugate of a real-valued function is of course itself.

In the Fourier representation,
the orthonormal separation functions are
$\psi_\alpha(r) = j_0(k_\alpha r)$
(compare eq.~[\ref{xik}]),
and summation over $\alpha$ signifies integration over
% Symmetric:
%$4\PI k_\alpha^2 \dd k_\alpha$.
% Asymmetric:
$4\PI k_\alpha^2 \dd k_\alpha /(2\PI)^3$.
The orthonormal pair functions are exponentials
$\phi_{ij}(\r_1,\r_2)
= {\rmn Sym}_{(ij)} \e^{\im \k_i.\r_1 + \im \k_j.\r_2}$,
and summation over $ij$ signifies integration over
% Symmetric:
%$\ddd k_i \ddd k_j$,
% Asymmetric:
$\ddd k_i \ddd k_j /(2\PI)^6$.
In Fourier space,
raising an index $i$, which means take the Hermitian conjugate with
respect to $\k_i$, is equivalent to replacing $\k_i \rightarrow -\k_i$,
for functions which are real-valued in their real space representation,
as is true for all functions considered in this paper.

Let $D^\alpha_{ij}$ denote the delta-function
$\delta_D (|\r_i-\r_j|-r_\alpha)$ in a general representation.
Explicitly,
with respect to arbitrary orthonormal bases
$\psi_\alpha(r)$ of separation functions
and $\phi_{ij}(\r_1,\r_2)$ of pair functions,
the matrix $D^\alpha_{ij}$ is
\ba
  D^\alpha_{ij}
  \!\!\! &=& \!\!\!
    \int \delta_D(r_{12} - r) \psi^\alpha(r) \phi_{ij}(\r_1,\r_2)
    \, 4\PI r^2 \dd r \, \ddd r_1 \ddd r_2
  \nn
  &=& \!\!\!
    \int \phi_{ij}(\r_1,\r_2) \psi^\alpha(r_{12})
    \, \ddd r_1 \ddd r_2
  \ .
\label{Dp}
\ea
Then in general the convention~(\ref{reduce}) of replacing a pair
index $ij$ with a separation index $\alpha$
is equivalent to contracting with the matrix $D^\alpha_{ij}$
\be
\label{reduceD}
  W^\alpha \equiv D^\alpha_{ij} W^{ij}
  \ .
\ee
As already indicated above,
in the real space representation $D^\alpha_{ij}$ is
\be
\label{D}
  D^\alpha_{ij} = \delta_D (|\r_i-\r_j|-r_\alpha)
  \ .
\ee
In the Fourier representation $D^\alpha_{ij}$ is
\be
\label{Dk}
  D^\alpha_{ij} =
% Symmetric:
%    (2\PI)^{3/2} \delta_D(\k_i+\k_j) \delta_D(k_i-k_\alpha)
% Asymmetric:
    (2\PI)^6 \delta_D(\k_i+\k_j) \delta_D(k_i-k_\alpha)
  \ .
\ee
The second delta-function in equation~(\ref{Dk}) is normalized so
% Symmetric and Asymmetric:
$\int \delta_D (k_i-k_\alpha) 4\PI k_\alpha^2 \dd k_\alpha = 1$.
Thus in Fourier space,
if $W^{ij} = W(-\k_i,-\k_j)$, then
% Symmetric:
%$W^\alpha = (2\PI)^{3/2} \int W(-k_\alpha\n,k_\alpha\n) \dd o/(4\PI)$,
% Asymmetric:
$W^\alpha = \int W(-k_\alpha\n,k_\alpha\n) \dd o/(4\PI)$,
where the integral over solid angle $\dd o$ is over all unit directions $\n$.
% Is following correct?
%In spherical harmonic basis
%$D^\alpha_{ij} =
%\delta_{\el_\alpha \el_i} \delta_{\el_\alpha \el_j} \delta_{m_i,-m_j}$.

\subsection{Minimum variance pair weighting}
\label{minvarpr}

Consider the problem of measuring the power spectrum $\xi_\alpha$
windowed over some arbitrary kernel function $\G^\alpha$
\be
\label{Gxi}
  \tilde\xi \equiv \G^\alpha \xi_\alpha
  \ .
\ee
Equation~(\ref{Gxi}) is equation~(\ref{Gxik}) in abbreviated,
representation-independent form.
The windowed power spectrum $\tilde\xi$ can be estimated by
an estimator $\hat\xi$
which is a weighted sum over pairs $\delta_i \delta_j$
of overdensities in the survey
\be
\label{Gxiest}
  \hat\xi = W^{ij} \delta_i \delta_j
\ee
where $W^{ij}$ is any a priori pair window constrained to satisfy
\be
\label{constraint}
  W^\alpha = \G^\alpha
\ee
and which vanishes along the diagonal in real space
\be
\label{WNP}
  W(\r,\r) = 0
  \ .
\ee
Equation~(\ref{Gxiest}) is equation~(\ref{xiest}) in abbreviated,
representation-independent form.
In equation~(\ref{constraint}),
$W^\alpha \equiv D^\alpha_{ij} W^{ij}$
is the representation-independent
symbol for $W^{ij}$ integrated over all pairs $ij$
at separation $\alpha$ in the survey,
in accordance with the convention~(\ref{reduceD}).
The condition that the window $W^{ij}$ be chosen a priori,
independently of knowledge of the densities $\delta_i$
(as is done below),
ensures that the window will be uncorrelated with the density,
and therefore that the estimator $\hat\xi$ will be unbiassed.

The estimate $\hat\xi$, equation~(\ref{Gxiest}),
of $\tilde\xi$ is valid because it is being assumed,
equation~(\ref{XiNP}),
that the expectation value
$\langle \delta(\r_i) \delta(\r_j) \rangle$
of overdensities in real space at points $\r_i \neq \r_j$
is equal to the cosmic variance $\xi(r_\alpha)$
at separation $|\r_i-\r_j| = r_\alpha$.
The condition~(\ref{WNP}) that the pair window
$W(\r_i,\r_j)$ in real space vanishes at $\r_i = \r_j$
ensures that the Poisson noise term in the survey covariance at
$\r_i = \r_j$ is excluded.
Since the estimator~(\ref{Gxiest}) is valid in the real space representation,
it follows immediately that it is valid in any representation.

Imposing the condition~(\ref{WNP}) that the pair window vanishes
along the diagonal in real space, $W(\r,\r) = 0$,
is equivalent to excluding from the computation of
$W^{ij} \delta_i \delta_j$
self-pairs of galaxies,
in which the pair consists of a galaxy and itself.
Alternatively,
it may be computationally convenient to include the Poisson contribution to
$W^{ij} \delta_i \delta_j$,
by including self-pairs of galaxies,
and to subtract the self-pair contribution as a subsequent step.
Such a subtraction is always exact.

The minimum variance pair weighting is that which minimizes the
expected variance of the estimator~(\ref{Gxiest}),
\be
\label{dxi}
  \langle \Delta\hat\xi^2 \rangle =
    W^{ij} W^{kl} \langle \Delta \XI_{ij} \Delta \XI_{kl} \rangle
  \ ,
\ee
subject to the constraints~(\ref{constraint}) and (\ref{WNP}).
The condition~(\ref{WNP}) on the pair window $W^{ij}$ can be discarded
provided that the covariance
$\langle \Delta \XI_{ij} \Delta \XI_{kl} \rangle$
used in equation~(\ref{dxi})
is taken to be the covariance~(\ref{XiXiNP}),
in which the terms with $\r_i = \r_j$ or $\r_k = \r_l$
present in the full covariance~(\ref{XiXi}) are excluded.
The constraints~(\ref{constraint}) can be imposed by
introducing Lagrange multipliers $\lambda_\alpha$,
and by minimizing the function
\be
\label{function}
  W^{ij} W^{kl} \langle \Delta \XI_{ij} \Delta \XI_{kl} \rangle
    - 2 \lambda_\alpha ( W^\alpha - \G^\alpha )
\ee
with respect to the weights $W^{ij}$ and the multipliers $\lambda_\alpha$.
Setting the partial derivatives of the function~(\ref{function})
with respect to the multipliers $\lambda_\alpha$ equal to zero
recovers the constraints~(\ref{constraint}),
while setting the partial derivatives
with respect to the weights $W^{kl}$ equal to zero implies
\be
\label{dfdW}
  W^{ij} \langle \Delta \XI_{ij} \Delta \XI_{kl} \rangle
    - \lambda_\alpha D^\alpha_{kl} = 0
  \ .
\ee
Equation~(\ref{dfdW}) implies
\be
\label{dfdW2}
  W^{ij} = T^{ij \alpha} \lambda_\alpha
\ee
where $T^{ijkl}$ is defined to be the inverse of the covariance matrix
$\langle \Delta \XI_{ij} \Delta \XI_{kl} \rangle$,
\be
\label{T}
  T^{ijkl} \equiv \langle \Delta \XI \Delta \XI \rangle^{-1 ijkl}
\ee
[so that
$\langle \Delta \XI_{ij} \Delta \XI_{mn} \rangle T^{mnkl}
= {\rmn Sym}_{(kl)} \delta_{Di}^{\ \ k} \delta_{Dj}^{\ \ l}$]
and
\be
\label{Ta}
  T^{ij \alpha} \equiv T^{ijkl} D^\alpha_{kl}
\ee
is the integral of this matrix over all pairs $kl$
separated by $\alpha$,
in accordance with the convention~(\ref{reduceD}).
For Gaussian fluctuations, equation~(\ref{XiXig}),
the inverse $T^{ijkl}$ takes the simplified form
\be
\label{Tg}
  T^{ijkl}
    = \frac{1}{2} \Sym{(kl)} \XI^{-1 ik} \XI^{-1 jl}
  \ ,
\ee
but the analysis here is not restricted to the Gaussian case.
The Lagrange multipliers $\lambda_\alpha$ can be eliminated
by summing equation~(\ref{dfdW2}) over pairs $ij$ at fixed separation
$\beta$
(that is, by applying the operator $D^\beta_{ij}$),
and imposing the constraints~(\ref{constraint}):
\be
  W^\beta = T^{\beta \alpha} \lambda_\alpha = \G^\beta
  \ ,
\ee
which shows $\lambda_\alpha = T_{\alpha \beta}^{-1} \G^\beta$.
Here again $T^{\alpha \beta}$ signifies $T^{ijkl}$
integrated over all pairs $ij$ separated by $\alpha$,
and all pairs $kl$ separated by $\beta$,
according to the convention~(\ref{reduceD}):
\be
\label{Tab}
  T^{\alpha \beta} \equiv D^\alpha_{ij} T^{ijkl} D^\beta_{kl}
  \ ,
\ee
and $T_{\alpha \beta}^{-1}$ denotes the inverse of $T^{\alpha \beta}$.
Thus the minimum variance pair weighting~(\ref{dfdW2}) is
\be
\label{minvar}
  W^{ij} = T^{ij \alpha} T_{\alpha \beta}^{-1} \G^\beta
  \ .
\ee

Equation~(\ref{minvar}) gives that pair weighting $W^{ij}$ which minimizes
the expected variance of the power spectrum windowed over some arbitrary
kernel $G$, equation~(\ref{Gxi}),
amongst estimators~(\ref{Gxiest})
quadratic in the observed overdensities $\delta$.

\section{Fisher Matrix}
\label{fisher}

Consider the expected covariance between estimates
$\hat\xi$ and $\hat\xi'$ of power spectra
$\tilde\xi = \G^\alpha \xi_\alpha$
and
$\tilde\xi' = \G'^\alpha \xi_\alpha$
windowed through kernels $\G^\alpha$ and $\G'^\alpha$.
If both estimates $\hat\xi$ and $\hat\xi'$
are made using minimum variance weightings
$W^{ij}$ and $W'^{ij}$,
then a short calculation from equation~(\ref{minvar}) shows that
the expected covariance
$\langle \Delta\hat\xi \Delta\hat\xi' \rangle$
of the estimates is
\be
\label{xixi}
  \langle \Delta\hat\xi \Delta\hat\xi' \rangle
    = W^{ij} W'^{kl} \langle \Delta\XI_{ij} \Delta\XI_{kl} \rangle
    = T_{\alpha \beta}^{-1} \G^\alpha \G'^\beta
  \ .
\ee
Equation~(\ref{xixi}) shows that the expected covariance matrix
$\langle \Delta\hat\xi_\alpha \Delta\hat\xi_\beta \rangle$
between minimum variance estimates
$\hat\xi_\alpha$ and $\hat\xi_\beta$
of power spectra is
\be
\label{xiaxibT}
  \langle \Delta\hat\xi_\alpha \Delta\hat\xi_\beta \rangle
    = T_{\alpha \beta}^{-1}
  \ .
\ee
The important matrix $T^{\alpha \beta}$, given by equation~(\ref{Tab}),
can be recognized as the Fisher information matrix
(Tegmark et al.\ 1997, \S2)
for the case where the parameter set to be measured is the power spectrum,
and the only source of noise is Poisson sampling noise,
as set forth in \S\ref{prior}.

Strictly, the Fisher matrix is defined as the matrix of second derivatives
of minus the log-likelihood function with respect to the parameters.
However,
the central limit theorem ensures that
estimates of quantities become Gaussianly distributed about their true values
in the limit of a sufficiently large survey,
so that the log-likelihood function is quadratic about its maximum.
Correctly, it is only in this limit of a large survey
that the Fisher matrix equals the inverse of the covariance matrix.
Here I tacitly assume that the survey at hand is large enough that
the central limit theorem applies.
It is to be noted that the statement of the central limit theorem,
that an estimate is asymptotically Gaussianly distributed about its true value,
is entirely distinct from the question of whether or not the density
fluctuations themselves form a multivariate Gaussian distribution.
The central limit theorem applies to non-Gaussian as well as Gaussian
density fields.

Some of the power of Fisher matrix $T^{\alpha \beta}$
will become apparent in Paper~2.

\section{Redshift Distortions}
\label{redshift}

The analysis hitherto addressed the problem of determining
the power spectrum from unredshifted data.
In redshift space however,
peculiar velocities of galaxies along the line of sight
distort the pattern of clustering.
In this section
I show how the results obtained so far extend to the
case of measuring the unredshifted power spectrum from redshift data,
at least for redshift distortions in the linear regime.
I show that the minimum variance pair window differs from the
unredshifted case, equation~(\ref{Ws}),
but that the Fisher matrix of the unredshifted power spectrum remains
unchanged.
The minimum variance window depends on the linear growth rate parameter
$\ff \approx \Omega^{0.6}/b$,
which is taken here to be a prior parameter.
I plan to address the issue of measuring $\ff$ itself in a subsequent paper.

Let a superscript $(s)$ denote quantities measured in redshift space.
For fluctuations in the linear regime, the overdensity $\delta^{(s)}$ in
redshift space is linearly related to the overdensity $\delta$ in real space by
\be
\label{deltas}
  \delta^{(s)} = S \delta
\ee
where $S$ is the linear redshift distortion operator,
which in the real space representation is
(Hamilton \& Culhane 1996, equation [12])
\be
\label{S}
  S =
   1
   + \ff \left( {\partial^2 \over \partial r^2}
   + {\alpha (\r) \partial \over r \partial r} \right) \nabla^{-2}
\ee
with $\ff \approx \Omega^{0.6}/b$ the linear growth rate parameter,
and $\alpha(\r)$ the logarithmic slope of $r^2$ times the selection
function $\Phi(\r)$ at depth $r$
\be
\label{alpha}
  \alpha(\r) \equiv {\partial \ln r^2 \Phi(\r) \over \partial \ln r}
  \ .
\ee

Consider a weighted sum of products of overdensities in redshift space
$W^{(s) ij} \delta^{(s)}_i \delta^{(s)}_j$.
According to the relation~(\ref{deltas}) between $\delta^{(s)}$ and $\delta$,
this weighted sum is
(here $S_i^{\ j}$ is the redshift distortion operator~(\ref{S})
in a general representation)
\be
\label{Wdd}
  W^{(s) ij} \delta^{(s)}_i \delta^{(s)}_j =
    W^{(s) ij} S_i^{\ k} S_j^{\ l} \delta_k \delta_l
    = W^{kl} \delta_k \delta_l
\ee
where the last equation can be regarded as defining a real space pair window
$W^{ij}$ equivalent to the redshift space pair window $W^{(s) ij}$.
Equation~(\ref{Wdd})
shows that the real and redshift space windows are related by
\be
\label{W}
  W^{kl}
    = W^{(s) ij} S_i^{\ k} S_j^{\ l}
    = S^{\dagger k}_{\ \ i} S^{\dagger l}_{\ \ j} W^{(s) ij}
\ee
where $S^{\dagger}$ is the Hermitian conjugate of the distortion operator $S$,
which in the real space representation is
\be
\label{Sdag}
  S^\dagger =
    1 + \ff \nabla^{-2} r^{-2} {\partial \over \partial r}
    \left( {\partial \over \partial r} - {\alpha(\r) \over r} \right) r^2
  \ .
\ee
Equation~(\ref{W}) inverts to give the redshift space pair window $W^{(s) ij}$
in terms of its unredshifted counterpart $W^{ij}$
\be
\label{Ws}
  W^{(s) ij} =
    S^{\dagger -1 i}_{\ \ \ \ \ k}
    S^{\dagger -1 j}_{\ \ \ \ \ l}
    W^{kl}
\ee
where $S^{\dagger -1}$ is the inverse, i.e.\ the Green's function,
of the Hermitian conjugate of the distortion operator $S$.
This inverse is known explicitly in cases where $S$ can be diagonalized,
which include the plane-parallel approximation (Kaiser 1987),
where $S$ is diagonalized in the Fourier representation,
and cases where $\alpha(\r)$, equation~(\ref{alpha}), is a constant,
for which $S$ is diagonalized in the representation of
logarithmic spherical waves
(Hamilton \& Culhane 1996, eq.~[50];
note that the Hermitian conjugate of $\eta$,
the eigenvalue of $- \partial/\partial\ln r$,
in this equation is $\eta^\dagger = 3 - \eta$;
see also Tegmark \& Bromley 1995,
who give an expression for the Green's function $S^{-1}$
for the particular case $\alpha(\r) = 2$).

The minimum variance pair window $W^{ij}$ in real space was
derived in \S\ref{minvariance}, equation~(\ref{minvar}).
The minimum variance pair window $W^{(s) ij}$ in redshift space
is then given in terms of the minimum variance window $W^{ij}$
by equation~(\ref{Ws}).
This is true because minimizing $W^{(s) ij} \delta^{(s)}_i \delta^{(s)}_j$
is equivalent to minimizing $W^{ij} \delta_i \delta_j$,
the two expressions being equal by definition~(\ref{Wdd}).

Although the minimum variance pair window differs between real
and redshift space,
the Fisher matrix of the unredshifted power spectrum is itself unchanged.
This is because it is being assumed that the quantity being estimated
is the unredshifted power spectrum,
whose expected covariance matrix
$\langle \Delta\hat\xi_\alpha \Delta\hat\xi_\beta \rangle$
depends on the survey geometry and the cosmic covariance,
but is independent of whether the data lie in real or redshift space.
A pertinent consideration here is that,
as emphasized by Fisher, Scharf \& Lahav (1994),
the selection function $\Phi(\r)$ in a flux-limited survey
is a function of the true distance $r$ to a galaxy, not the redshift distance.

The same conclusion,
that the Fisher matrix of the unredshifted power spectrum is the same
whether the data lie in real or redshift space,
is found if the analysis in
\S\ref{minvariance}
is carried through in redshift space.
The redshift space versions of relevant quantities are as follows.
The covariance matrix
$\langle \Delta\XI^{(s)}_{ij} \Delta\XI^{(s)}_{kl} \rangle$
of the survey covariance matrix in redshift space is related to
its unredshifted counterpart
$\langle \Delta\XI_{ij} \Delta\XI_{kl} \rangle$
by
\be
\label{XisXis}
  \langle \Delta\XI^{(s)}_{ij} \Delta\XI^{(s)}_{kl} \rangle =
    S_i^{\,m} S_j^{\ n} S_k^{\ p} S_l^{\ q}
    \langle \Delta\XI_{mn} \Delta\XI_{pq} \rangle
  \ .
\ee
Similarly the inverse $T^{(s) ijkl}$ of this redshift covariance matrix
is related to its unredshifted counterpart $T^{ijkl}$ by
\ba
  T^{(s) ijkl} &\equiv&
    \langle \Delta \XI^{(s)} \Delta \XI^{(s)} \rangle^{-1 ijkl}
  \nn
  &=&
\label{Ts}
    T^{mnpq}
    S_{\ m}^{-1 i} S_{\ \,n}^{-1 j}
    S_{\ \,p}^{-1 k} S_{\ \,q}^{-1 l}
  \ .
\ea
The relation between the matrix $D^{(s) \alpha}_{\ \ \,ij}$ in redshift space
and its unredshifted counterpart $D^\alpha_{ij}$ is determined by the
requirement that
\be
\label{DW}
  D^\alpha_{kl} W^{kl}
    = D^\alpha_{kl} W^{(s) ij} S_i^{\ k} S_j^{\ l}
    = D^{(s) \alpha}_{\ \ \,ij} W^{(s) ij}
\ee
in which the first equality follows from the relation~(\ref{W})
between the pair windows $W^{ij}$ and $W^{(s) ij}$,
and the second equality can be regarded as defining $D^{(s) \alpha}_{\ \ \,ij}$.
Equation~(\ref{DW}) shows that
$D^{(s) \alpha}_{\ \ \,ij}$ is related to $D^\alpha_{ij}$ by
\be
\label{Ds}
  D^{(s) \alpha}_{\ \ \,ij}
    = D^\alpha_{kl} S_i^{\ k} S_j^{\ l}
  \ .
\ee
Finally,
the Fisher matrix $T^{(s) \alpha \beta}$ in redshift space is
\be
\label{Tsab}
  T^{(s) \alpha \beta} \equiv
    D^{(s) \alpha}_{\ \ \,ij} T^{(s) ijkl} D^{(s) \beta}_{\ \ \,kl}
    = T^{\alpha \beta}
\ee
in which the last equality follows from a short calculation from
equations~(\ref{Ts}) and (\ref{Ds}).
Thus the Fisher matrix $T^{\alpha \beta}$ is the same matrix
in both real and redshift space,
as claimed.

\section{The Classical Limit}
\label{classical}

\subsection{Gaussian Fluctuations}
\label{classicalg}

Feldman, Kaiser \& Peacock (1994; hereafter FKP)
derived a minimum variance pair window
$\propto [\xi(k) + \Phi(\r_i)^{-1}]^{-1} \discretionary{}{}{}
[\xi(k) + \Phi(\r_j)^{-1}]^{-1}$
for measuring the power spectrum $\xi(k)$,
valid for Gaussian fluctuations in the limit where
the selection function varies slowly compared to the wavelength,
which may be termed the classical limit.
In this Section it is shown how FKP's pair window,
equation~(\ref{Wc}),
emerges from the present analysis.
The classical solution will be used in the next Section, \S\ref{perturb},
as the starting point of a general perturbative expansion
for the minimum variance pair window and the Fisher matrix.

The survey covariance matrix $\XI_{ij}$, equation~(\ref{Xi}),
can be regarded as an operator which is the sum of two operators,
$\XI = \xi + \Phi^{-1}$,
the first of which,
the cosmic covariance
% Symmetric:
%$(2\PI)^{3/2} \delta_D(\k_i+\k_j) \xi(k_i)$,
% Asymmetric:
$(2\PI)^3 \delta_D(\k_i+\k_j) \xi(k_i)$,
is diagonal in Fourier space,
and the second of which,
the inverse selection function $\delta_D(\r_i-\r_j)\Phi(\r_i)^{-1}$,
is diagonal in real space.
The two terms do not commute in general,
but they do commute approximately,
$\xi(r_{ij}) \Phi(\r_j)^{-1} - \Phi(\r_i)^{-1} \xi(r_{ij}) \approx 0$,
in the classical limit where the wavelength is much shorter
than the scale over which the selection function varies.
In the classical limit,
wavelength and position can be measured simultaneously,
and the two terms in the survey covariance $\XI_{ij}$ can be diagonalized
simultaneously:
\be
\label{Xic}
  \XI_{ij} = \delta_{D ij} [\xi(k_i) + \Phi(\r_i)^{-1}]
  \ .
\ee
The eigenfunctions of $\XI_{ij}$ in this representation are wave packets
localized in both position and wavelength,
and $\delta_{D ij}$ here should be interpreted
as the unit matrix in this representation.
Equation~(\ref{Xic}) means that $\XI_{ij}$ acting on any function localized
around position $\r_i$ and wavelength $\k_i$ is equivalent to multiplication by
$\xi(k_i) + \Phi(\r_i)^{-1}$.
The inverse $\XI^{-1 ij}$ of the survey covariance
follows immediately from equation~(\ref{Xic}),
and is likewise diagonal when acting on functions localized in position
and wavelength
\be
\label{Xiinvc}
  \XI^{-1 ij} = {\delta_D^{ij} \over \xi(k_i) + \Phi(\r_i)^{-1}}
  \ .
\ee

For Gaussian fluctuations,
the covariance
$\langle \Delta\XI_{ij} \Delta\XI_{kl} \rangle$
of the survey covariance
is equal to
$\XI_{ik} \XI_{jl} + \XI_{il} \XI_{jk}$,
equation~(\ref{XiXig}),
and is therefore also diagonal
in the classical limit,
in a representation where
the eigenfunctions are products of pairs of wave-packets
localized in position and wavelength:
\ba
  \langle \Delta\XI_{ij} \Delta\XI_{kl} \rangle
  \!\!\! &=& \!\!\!
    ( \delta_{D ik} \delta_{D jl}
    + \delta_{D il} \delta_{D jk} )
  \nn
\label{XiXic}
  && \!\!\!
    \times \ 
% Symmetric:
%    [(2\PI)^{3/2} \xi(k_i) + \Phi(\r_i)^{-1}]
%    [(2\PI)^{3/2} \xi(k_j) + \Phi(\r_j)^{-1}]
% Asymmetric:
    [\xi(k_i) + \Phi(\r_i)^{-1}]
    [\xi(k_j) + \Phi(\r_j)^{-1}]
  \ .
\ea
This expression remains valid even when the wave-packets $i$ and $j$
are well-separated in position and/or wavelength:
the delta-functions in equation~(\ref{XiXic}) assert only that
wave-packet $i$ is the same as wave-packet $k$ (or $l$),
and that $j$ is the same as $l$ (or $k$).
The inverse
$T^{ijkl}$
of the covariance matrix~(\ref{XiXic}) is again diagonal
when acting on pair functions localized in position and wavelength
\be
\label{Tc}
  T^{ijkl} =
    {\delta_D^{ik} \delta_D^{jl}
    + \delta_D^{il} \delta_D^{jk}
    \over 4
% Symmetric:
%    [(2\PI)^{3/2} \xi(k_i) + \Phi(\r_i)^{-1}]
%    [(2\PI)^{3/2} \xi(k_j) + \Phi(\r_j)^{-1}]}
% Asymmetric:
    [\xi(k_i) + \Phi(\r_i)^{-1}]
    [\xi(k_j) + \Phi(\r_j)^{-1}]}
  \ .
\ee

The minimum variance pair window~(\ref{minvar}) derived in \S\ref{minvariance}
involves the quantities
$T^{ij \alpha} = T^{ijkl} D^\alpha_{kl}$
and $T^{\alpha \beta} = D^\alpha_{ij} T^{ijkl} D^\beta_{kl}$,
where $D^\alpha_{ij}$ is the operator~(\ref{Dp}) which
effectively integrates over pairs $ij$ at separation $\alpha$.
Now $T^{ijkl}$ above is diagonal provided that it acts on functions
which are localized in both position and wavelength.
This can be accomplished by choosing
$D^\alpha_{ij}$
in a mixed representation with $ij$ in real space and $\alpha$ in
Fourier space, in which case $D^\alpha_{ij}$ is a spherical Bessel function
\be
\label{Dc}
% Symmetric:
%  D^\alpha_{ij} = (2/\PI)^{1/2} j_0(k_\alpha r_{ij})
% Asymmetric:
  D^\alpha_{ij} = j_0(k_\alpha r_{ij})
\ee
with $r_{ij} \equiv |\r_i-\r_j|$.
Contracting the matrix $T^{ijkl}$ in equation~(\ref{Tc})
with $D^\alpha_{kl}$ from equation~(\ref{Dc}) replaces the
Dirac delta-functions in the numerator of $T^{ijkl}$
with $D^{\alpha ij}$,
and sets $k_i = k_j = k_\alpha$,
the latter being evident from the Fourier representation~(\ref{Dk}) of
$D^\alpha_{ij}$.
The resulting matrix $T^{ij \alpha}$ is, in the mixed representation,
\be
\label{Tac}
  T^{ij\alpha} =
% Symmetric:
%    {(2/\PI)^{1/2} j_0(k_\alpha r_{ij})
% Asymmetric:
    {j_0(k_\alpha r_{ij})
    \over 2
% Symmetric:
%    [(2\PI)^{3/2} \xi(k_\alpha) + \Phi(\r_i)^{-1}]
%    [(2\PI)^{3/2} \xi(k_\alpha) + \Phi(\r_j)^{-1}]}
% Asymmetric:
    [\xi(k_\alpha) + \Phi(\r_i)^{-1}]
    [\xi(k_\alpha) + \Phi(\r_j)^{-1}]}
  \ .
\ee
This is,
up to a normalization factor, the
FKP
pair window.
Operating again on $T^{ij\alpha}$ in equation~(\ref{Tac})
with $D^\beta_{ij}$ from equation~(\ref{Dc})
yields the Fisher matrix $T^{\alpha \beta}$
in the Fourier representation
\be
\label{Tabc}
  T^{\alpha \beta} =
% Symmetric:
%   {2 \over \PI}
% Asymmetric:
% nothing
    \int_0^\infty \!
    {j_0(k_\alpha r_{ij}) j_0(k_\beta r_{ij}) \, \ddd r_i \ddd r_j
    \over 2
% Symmetric:
%    [(2\PI)^{3/2} \xi(k_\alpha) + \Phi(\r_i)^{-1}]
%    [(2\PI)^{3/2} \xi(k_\alpha) + \Phi(\r_j)^{-1}]}
% Asymmetric:
    [\xi(k_\alpha) + \Phi(\r_i)^{-1}]
    [\xi(k_\alpha) + \Phi(\r_j)^{-1}]}
  \ .
\ee
In the same classical limit
that the selection function $\Phi$ varies slowly over many wavelengths,
the Fisher matrix $T^{\alpha \beta}$ is diagonal
in Fourier space
\be
\label{Tabd}
  T^{\alpha \beta} =
% Asymmetric:
    (2\PI)^3 \delta_D(k_\alpha-k_\beta)
    \int {\ddd r \over 2 [\xi(k_\alpha) + \Phi(\r)^{-1}]^2}
  \ .
\ee

From equations~(\ref{minvar}) and (\ref{Tac})
it follows that the minimum variance pair window $W^{ij}$
for measuring the power spectrum $\xi(k_\alpha)$ at wavenumber $k_\alpha$
in the classical limit is, in real space,
\be
\label{Wc}
  W(\r_i,\r_j) =
% Asymmetric:
    {\sigma_\alpha^2 \, j_0(k_\alpha r_{ij})
    \over 2
% Symmetric:
%    [(2\PI)^{3/2} \xi(k_\alpha) + \Phi(\r_i)^{-1}]
%    [(2\PI)^{3/2} \xi(k_\alpha) + \Phi(\r_j)^{-1}]}
% Asymmetric:
    [\xi(k_\alpha) + \Phi(\r_i)^{-1}]
    [\xi(k_\alpha) + \Phi(\r_j)^{-1}]}
\ee
where $\sigma_\alpha^2$ is the reciprocal of
the eigenvalue of the Fisher matrix $T^{\alpha \beta}$,
this eigenvalue being the coefficient of the identity matrix
$(2\PI)^3 \delta_D(k_\alpha-k_\beta)$ in equation~(\ref{Tabd}),
\be
\label{sigma}
  \sigma_\alpha^2 \equiv
    \left[
% Asymmetric:
    \int {\ddd r \over 2 [\xi(k_\alpha) + \Phi(\r)^{-1}]^2}
    \right]^{-1}
  \ .
\ee
Equation~(\ref{Wc}) gives the pair window~(\ref{Tac}) correctly normalized
to yield an estimate $\hat\xi(k_\alpha)$ of the power spectrum
at wavenumber $k_\alpha$
\be
\label{xiFKP}
  \hat\xi(k_\alpha) =
    \int W(\r_i,\r_j) \delta(\r_i) \delta(\r_j) \, \ddd r_i \ddd r_j
  \ .
\ee
Equations~(\ref{Wc}) and (\ref{xiFKP})
are precisely FKP's estimator of the power spectrum.

The expected covariance
$\langle \Delta\hat\xi(k_\alpha) \Delta\hat\xi(k_\beta) \rangle$
between estimates~(\ref{xiFKP}) of the power spectrum is,
according to equation~(\ref{xiaxibT}),
given by the inverse of the Fisher matrix $T^{\alpha \beta}$
in equation~(\ref{Tabd}),
\be
\label{xiaxibc}
  \langle \Delta\hat\xi(k_\alpha) \Delta\hat\xi(k_\beta) \rangle
    = (2\PI)^3 \delta_D(k_\alpha - k_\beta) \, \sigma_\alpha^2
  \ .
\ee
The delta-function here means that the variance diverges for
$k_\alpha = k_\beta$.
To resolve the divergence,
it is necessary to average estimates $\hat\xi(k_\alpha)$ of the power spectrum
over shells of finite thickness in $k$-space.
Of course the divergence is only an artefact of the classical limit:
in reality the delta-function in equation~(\ref{xiaxibc})
has a finite coherence width of the order of the inverse scale of the survey.
Thus it is necessary to average over shells broad enough to
ensure that estimates of the power spectrum in neighbouring shells
are effectively independent.
So let $\bar\xi(k_\alpha)$ denote the estimated power spectrum~(\ref{xiFKP})
averaged over a shell of volume $V_k$ about $k_\alpha$ which is thin
(hopefully) compared to the scale over which $\xi(k)$ varies,
yet broad compared to a coherence length
\be
\label{xibar}
  \bar\xi(k_\alpha) \equiv V_k^{-1} \int \hat\xi(k_\alpha) \, \dd V_k
\ee
where $\dd V_k \equiv 4\PI k_\alpha^2 \dd k_\alpha / (2\PI)^3$.
FKP call $V_k$ the coherence volume.
The variance of the shell-averaged power spectrum $\bar\xi(k_\alpha)$ is then
\be
\label{dxibar}
  \langle \Delta\bar\xi(k_\alpha)^2 \rangle =
    \sigma_\alpha^2 / V_k
\ee
which shows that the variance decreases inversely with the coherence volume
$V_k$, as expected for the variance of averages of independent quantities.

\subsection{Non-Gaussian fluctuations}
\label{classicalng}

Unfortunately,
the FKP approach does not generalize in an elegant way to the
case of non-Gaussian fluctuations.
The fundamental difficulty is that, whereas
in the Gaussian case
the cosmic and Poisson sampling terms in the survey covariance
commuted in the classical limit of slowingly varying selection function $\Phi$,
in the non-Gaussian case the terms no longer commute.
That is, there are three sets of terms in the survey covariance
$\langle \Delta\XI_{ij} \Delta\XI_{kl} \rangle$,
equation~(\ref{XiXiNP}),
one independent of the selection function $\Phi$ (the cosmic term),
one proportional to $\Phi^{-1}$,
and one proportional to $\Phi^{-2}$.
The coefficients of these terms do not generally commute,
for non-Gaussian fluctuations.
If the three coefficients did commute,
then it would be possible to find simultaneous eigenfunctions of the terms,
from which localized `pair-wave-packets' could be constructed,
and it would be possible to proceed in the much the same way as the
Gaussian case of the previous section, \S\ref{classicalg}.
It turns out that a scalar separation variable can always be defined
in the space of such eigenfunctions,
analogous to the scalar wavenumber $k$ in the Gaussian case,
and the Fisher matrix would then be diagonal, in the classical limit,
with respect to this scalar separation variable,
in the same way that the Fisher matrix~(\ref{Tabd}) is diagonal in
Fourier space, for Gaussian fluctuations in the classical limit.

As it is,
it appears that the best that can be done,
in the classical limit of slowly varying selection function $\Phi$,
is to diagonalize the survey covariance
$\langle \Delta\XI_{ij} \Delta\XI_{kl} \rangle$
locally, which certainly is feasible.
The disadvantage of this is that the eigenfunctions depend on the
local value of the selection function $\Phi$,
and in particular the scalar separation variable associated with the
eigenfunctions has a different meaning depending on the local value of $\Phi$.
Whilst it may be possible to make headway along these lines,
here I do not pursue the issue further.

\section{Perturbative Solution}
\label{perturb}

In this section I show how the matrix $T^{ijkl}$,
and hence the minimum variance pair window and the
Fisher matrix $T^{\alpha \beta}$,
can be evaluated perturbatively.
In the Gaussian case,
the classical solution (in effect, the FKP approximation) of \S\ref{classicalg}
provides a natural zeroth order approximation.
Indeed, in practical applications
the zeroth order solution
may be judged already to be adequate, as in Paper~2.
For non-Gaussian fluctuations,
a perturbation solution can still be developed,
although the absence of a natural zeroth order approximation makes
this solution, at least for the time being, less useful.

\subsection{Gaussian fluctuations}
\label{perturbg}

For Gaussian fluctuations,
the inverse $T^{ijkl}$
of the covariance matrix
$\langle \Delta\XI_{ij} \Delta\XI_{kl} \rangle$
simplifies to a symmetrized product of the inverse
$\XI^{-1}$ of the survey covariance $\XI$, equation~(\ref{Tg}).
Thus to develop a perturbation series for $T$,
it suffices to develop a perturbation series for $\XI^{-1}$.

Suppose that $\up{0}{\XI^{-1 ij}}$
is a zeroth order approximation to the inverse of $\XI_{ij}$.
Multiplying the two matrices together yields
(with indices dropped for brevity)
\be
\label{epsilondefg}
  \XI \up{0}{\XI^{-1}} = 1 + \epsilon
\ee
where $\epsilon_i^{\ j}$ is a supposedly small matrix.
The exact inverse $\XI^{-1}$ is then given by
the perturbation series
\ba
  \XI^{-1}
    &=& \up{0}{\XI^{-1}} + \up{1}{\XI^{-1}} + \up{2}{\XI^{-1}} + \cdots
  \nn
    &=& \up{0}{\XI^{-1}} (1 - \epsilon + \epsilon^2 - \cdots)
\label{seriesg}
\ea
with
\be
\label{Xin}
  \up{n}{\XI^{-1}} = \up{0}{\XI^{-1}} (-\epsilon)^n
\ee
provided that the series converges.
In practice,
it may be that the series~(\ref{seriesg}) converges for some elements
of the matrix, but not for others.
If $\epsilon$ were diagonal for example,
then the series for any particular diagonal element $\epsilon_{i}^{\ i}$
would converge or diverge depending on whether
the absolute value of that element were less than or more than one.
In the actual case,
the matrix $\epsilon$ is only approximately diagonal,
so that the convergence of apparently converging elements may be asymptotic
-- the series converges up to a certain point, but thereafter diverges,
because of gradual mixing in of divergent parts of the matrix.
Below, I {\em assume}\/ that if the first few terms of the series
for a particular element of the matrix $\XI^{-1}$ are appearing to converge,
then they are converging to the true value of that element.
The complete inverse matrix $\XI^{-1}$ can then be built up
by combining pieces from several different choices of the initial guess
$\up{0}{\XI^{-1}}$.

To construct the zeroth order inverse $\up{0}{\XI^{-1 ij}}$,
start with the classical approximation~(\ref{Xiinvc}).
To be specific,
interpret the delta-function $\delta_D^{ij}$ in the numerator
as lying in real space,
and fix the wavenumber $k$ in the denominator
to be some constant $k_0$,
which will be adjusted later so as to make the first and higher
order corrections small.
The choice of $k_0$ is discussed at the end of this subsection~\ref{perturbg},
and further in \S\ref{approximate};
clearly one will want ultimately to choose $k_0$ to be close to the particular
wavenumber(s) at which one is trying to estimate the
the power spectrum, or its inverse variance, the Fisher matrix.
Thus the zeroth order inverse $\up{0}{\XI^{-1}_{ij}}$ is taken to be
(note that for functions which are real-valued in real space,
the value of a quantity with a raised index is the same as
that with a lowered index,
in the real space representation)
\be
\label{Xi0g}
  \up{0}{\XI^{-1}}(\r_i,\r_j) =
    \delta_D(\r_{ij}) U(\r_i)
\ee
where
$U(\r)$ is defined by
\be
\label{U}
  U(\r) \equiv
    {\Phi(\r) \over 1 + \xi(k_0) \Phi(\r)}
  \ ,
\ee
which may be regarded as the survey window, weighted in a certain way
(the FKP way, in fact).

According to the argument accompanying equation~(\ref{Xiinvc}),
the approximation~(\ref{Xi0g}) to the matrix $\XI^{-1}_{ij}$
only has limited validity,
namely it is valid when acting on functions
which are sufficiently localized in Fourier space about wavenumbers
$k \approx k_0$ that $\xi(k) \approx \xi(k_0)$
(the approximation~(\ref{Xi0g}) would have general validity
only if the power spectrum were nearly that of shot noise,
$\xi(k)$ = constant).
Thus it makes sense to pass into Fourier space,
and to develop the perturbation expansion there.
In the Fourier representation, the zeroth order inverse
$\up{0}{\XI^{-1}_{ij}}$
is
(it is useful to recall that for functions which are real-valued
in real space, as are all the functions in this section,
raising or lowering an index $i$ in Fourier space
is equivalent to changing $\k_i \rightarrow -\k_i$;
I adhere to the convention that a lowered index $i$ signifies $+\k_i$,
while a raised index signifies $-\k_i$)
\be
\label{Xi0k}
  \up{0}{\XI^{-1}}(\k_i,\k_j) = U(\k_i+\k_j)
\ee
where
$U(\k) = \int U(\r) \e^{\im \k.\r} \ddd r$
is the Fourier transform of $U(\r)$, the constant $k_0$ being held fixed
as the Fourier transform is taken.
In Fourier space,
the survey window $U(\k)$ is expected to be a compact ball about $\k = 0$,
of width $\sim 1/r$ where $r$ is the depth of the survey.
Thus $\up{0}{\XI^{-1}_{ij}}$ is expected to be near diagonal in Fourier space,
with only elements $i \approx j$ appreciably nonzero.

Multiplying the survey covariance $\XI_{ij}$, equation~(\ref{Xi}),
by the approximation~(\ref{Xi0g}) to its inverse,
and Fourier transforming,
yields the matrix $\epsilon_{ij}$ defined by equation~(\ref{epsilondefg})
\be
\label{epsilong}
  \epsilon(\k_i,\k_j) = U(\k_i+\k_j) [\xi(k_i) - \xi(k_0)]
  \ .
\ee
Again, the expectation that $U(\k)$ is concentrated about $\k = 0$ means that
the matrix $\epsilon_{ij}$ should be near diagonal in Fourier space,
with only elements $i \approx j$ appreciably nonzero.
Equation~(\ref{epsilong}) shows that the matrix $\epsilon$
is the product of the potentially small factor
$\xi(k_i) - \xi(k_0)$
times a factor $U$
which is no larger than $1/\xi(k_0)$, equation~(\ref{U}).
Thus it is expected that the series~(\ref{seriesg}) in $\epsilon$
should converge for wavenumbers $k$ satisfying
\be
  \left| {\xi(k) - \xi(k_0) \over \xi(k_0)} \right| \la 1
  \ .
\ee

The higher order perturbations
$\up{n}{\XI^{-1}}$
follow from equation~(\ref{Xin}) with
$\up{0}{\XI^{-1}}$ given by~(\ref{Xi0k})
and $\epsilon$ given by~(\ref{epsilong}).
The first order perturbation $\up{1}{\XI^{-1}_{ij}}$ is
\be
  \up{1}{\XI^{-1}}(\k_i,\k_j) =
    - \!\int\!
    U(\k_i-\k) U(\k_j+\k)
    [\xi(k)-\xi(k_0)]
    {\ddd k \over (2\PI)^3}
\ee
while the second order perturbation $\up{2}{\XI^{-1}_{ij}}$ is
\ba
  \up{2}{\XI^{-1}}(\k_i,\k_j)
  \!\!\! &=& \!\!\!
    \int
    U(\k_i-\k_1) U(\k_j-\k_2) U(\k_1+\k_2)
  \nn
  && \!\!\!
    [\xi(k_1)-\xi(k_0)] [\xi(k_2)-\xi(k_0)]
    {\ddd k_1 \ddd k_2 \over (2\PI)^6}
  \ .
\ea

The perturbation expansion of the matrix $T$
\be
\label{seriesTg}
  T = \up{0}{T} + \up{1}{T} + \up{2}{T} + \cdots
\ee
in orders of $\xi(k) - \xi(k_0)$ follows from
the Gaussian expression~(\ref{Tg}) for $T$ in terms of $\XI^{-1}$,
and the perturbation expansion~(\ref{seriesg}) of $\XI^{-1}$.
In the real space representation,
the zeroth order inverse $\up{0}{T_{ijkl}}$ is
\be
\label{T0g}
  \up{0}{T}(\r_i,\r_j,\r_k,\r_l) =
    \frac{1}{2} \Sym{(kl)}
    \delta_D(\r_{ik}) \delta_D(\r_{jl})
    U(\r_i) U(\r_j)
\ee
while in the Fourier representation $\up{0}{T_{ijkl}}$ is
\be
\label{T0kg}
  \up{0}{T}(\k_i,\k_j,\k_k,\k_l) =
    \frac{1}{2} \Sym{(kl)} U(\k_i+\k_k) U(\k_j+\k_l)
  \ .
\ee
The first order correction $\up{1}{T_{ijkl}}$ is
\be
\label{T1kg}
  \up{1}{T}(\k_i,\k_j,\k_k,\k_l) =
    \Sym{(ij)(kl)}
    U(\k_i+\k_k) \up{1}{\XI^{-1}}(\k_j,\k_l)
\ee
while the second order correction $\up{2}{T_{ijkl}}$ is
\ba
  \up{2}{T}(\k_i,\k_j,\k_k,\k_l)
  \!\!\! &=& \!\!\!
    \Sym{(ij)(kl)} \Bigl[
    \frac{1}{2} \up{1}{\XI^{-1}}(\k_i,\k_k) \up{1}{\XI^{-1}}(\k_j,\k_l)
  \nn
\label{T2kg}
  &&
    \mbox{}
    + U(\k_i+\k_k) \up{2}{\XI^{-1}}(\k_j,\k_l)
    \Bigr]
  \ .
\ea

The minimum variance pair window~(\ref{minvar})
involves the quantities
$T^{ij \alpha} = T^{ijkl} D^\alpha_{kl}$
and the Fisher matrix $T^{\alpha \beta} = D^\alpha_{ij} T^{ijkl} D^\beta_{kl}$.
The perturbation expansions of $T^{ij \alpha}$ and $T^{\alpha \beta}$
follow straightforwardly from the perturbation expansion of
$T^{ijkl}$ above, equations~(\ref{T0kg})-(\ref{T2kg}),
the matrix $D^\alpha_{ij}$ being given in the Fourier representation by
equation~(\ref{Dk}).
The zeroth order contribution $\up{0}{T^{ij \alpha}}$ to $T^{ij \alpha}$ is
\be
\label{T0ag}
  \up{0}{T^{ij \alpha}} =
    \frac{1}{2} \int
    U(-\k_i-\k_\alpha) U(-\k_j+\k_\alpha)
    \, \dd o_\alpha / (4\PI)
  \ ,
\ee
while the first and second order perturbations are
\be
\label{T1ag}
  \up{1}{T^{ij \alpha}} =
    \Sym{(ij)} \int U(-\k_i-\k_\alpha) \up{1}{\XI^{-1}}(-\k_j,\k_\alpha)
    \, \dd o_\alpha / (4\PI)
\ee
and
\ba
  \up{2}{T^{ij \alpha}}
  \!\!\! &=& \!\!\!
    \Sym{(ij)} \int \Bigl[
    \frac{1}{2}
    \up{1}{\XI^{-1}}(-\k_i,-\k_\alpha)
    \up{1}{\XI^{-1}}(-\k_j,\k_\alpha)
  \nn
\label{T2ag}
  && \!\!\!
    \mbox{}
    + U(-\k_i-\k_\alpha) \up{2}{\XI^{-1}}(-\k_j,\k_\alpha)
    \Bigr]
    \, \dd o_\alpha / (4\PI)
  \ .
\ea

For the Fisher matrix $T^{\alpha \beta}$,
the zeroth order term is
\be
\label{T0abg}
  \up{0}{T^{\alpha \beta}} =
    \frac{1}{2} \int \bigl| U(\k_\alpha+\k_\beta) \bigr|^2
    \, \dd o_\alpha \dd o_\beta / (4\PI)^2
\ee
while the first and second order terms are
\be
\label{T1abg}
  \up{1}{T^{\alpha \beta}} =
    - \int
    U(-\k_\alpha-\k_\beta) \up{1}{\XI^{-1}}(\k_\alpha,\k_\beta)
    \, \dd o_\alpha \dd o_\beta / (4\PI)^2
\ee
and
\ba
  \up{2}{T^{\alpha \beta}}
  \!\!\! &=& \!\!\!
    \int \Bigl[
    \frac{1}{2} \bigl| \up{1}{\XI^{-1}}(\k_\alpha,\k_\beta) \bigr|^2
  \nn
\label{T2abg}
  && \!\!\!
    \mbox{}
    + U(-\k_\alpha-\k_\beta) \up{2}{\XI^{-1}}(\k_\alpha,\k_\beta)
    \Bigr] \, \dd o_\alpha \dd o_\beta / (4\PI)^2
  \ .
\ea

The time has come to choose the adjustable constant $k_0$
in the definition~(\ref{U}) of the window $U$.
Suppose first that it is the Fisher matrix by itself which is of interest.
This is the case for example in Paper~2,
where the Fisher matrix is used to design sets of kernels for measuring
the power spectrum.
The zeroth order approximation
$\up{0}{T^{\alpha \beta}}$
to the Fisher matrix in Fourier space is given by equation~(\ref{T0abg}).
The presumed narrowness of the window $U(\k)$ about $\k = 0$
ensures that $k_\alpha \approx k_\beta$.
The perturbations $\up{1}{T^{\alpha \beta}}$ and $\up{2}{T^{\alpha \beta}}$
can then be made small if $k_0$ is chosen close to $k_\alpha$ and $k_\beta$.
The strategy adopted in Paper~2 is to set
$k_0 = (k_\alpha + k_\beta)/2$,
and to retain only the zeroth order term $\up{0}{T^{\alpha \beta}}$
of the Fisher matrix.
It would also be possible to choose $k_0$ more crudely,
if one were willing to retain additional perturbation terms.

Once a particular kernel or set of kernels has been chosen,
perhaps along the lines described in Paper~2,
it remains to estimate the power spectrum windowed through such kernels.
The minimum variance pair window~(\ref{minvar}) for estimating
the power spectrum windowed through a specified kernel $\G^\alpha$
involves both $T^{ij \alpha}$ and the Fisher matrix $T^{\alpha \beta}$.
If the kernel concerned is narrow about some wavenumber,
then it would be natural to set $k_0$ equal to that wavenumber
in approximating $T^{ij \alpha}$ and $T^{\alpha \beta}$.
The problem of how to apply an approximate pair window
is discussed further in \S\ref{approximate}.

\subsection{Non-Gaussian fluctuations}
\label{perturbng}

The perturbation expansion of the inverse $T^{ijkl}$ of the survey covariance
can be carried out for non-Gaussian fluctuations,
at least in principle,
in much the same way as in the Gaussian case.

Suppose that $\up{0}{T^{ijkl}}$ is an approximate inverse
of the covariance matrix
$\langle \Delta\XI_{ij} \Delta\XI_{kl} \rangle$,
where the latter now includes the non-Gaussian terms,
equation~(\ref{XiXiNP}).
Multiplying the two matrices together yields
\be
\label{epsilondef}
  \langle \Delta\XI \Delta\XI \rangle \, \up{0}{T} = 1 + \varepsilon
\ee
where $\varepsilon_{ij}^{\ \ kl}$ is a supposedly small matrix
(note the slightly different font for $\varepsilon$ here versus $\epsilon$
in~[\ref{epsilondefg}]).
The exact inverse $T^{ijkl}$ is then given by the perturbation expansion
\be
\label{series}
  T \equiv \langle \Delta\XI \Delta\XI \rangle^{-1} =
    \up{0}{T} (1 - \varepsilon + \varepsilon^2 - \cdots)
\ee
provided that the series converges.

Unfortunately,
it less clear here what to choose for the zeroth order approximation,
since, as already discussed in \S\ref{classicalng},
there does not appear to be an elegant non-Gaussian generalization
of the classical FKP approximation in the Gaussian case.
Lacking such a generalization,
I do not pursue the matter further here.

\section{Application of an Approximate Pair Window}
\label{approximate}

The approximations to the matrix $T^{ij \alpha}$ and the Fisher matrix
$T^{\alpha \beta}$ suggested in \S\ref{perturb}
yield approximations to the minimum variance window~(\ref{minvar})
derived in \S\ref{minvariance}.
The approximate pair window is of course not the same as the true
minimum variance pair window.
In applying an approximate pair window,
one should be careful about three points.
Firstly,
whatever pair window is adopted,
it should at least give an unbiassed estimate of the
quantity one wishes to measure
-- that is, the expectation value of the estimator should equal the
quantity it is desired to measure,
even if the variance in the estimator is not minimal.
Secondly,
the estimated uncertainty in the estimate
should be the uncertainty in the actual estimator used,
not the uncertainty in the minimum variance estimator.
Thirdly,
one should be aware that the minimum variance estimator
for data in redshift space
is not the same as the minimum variance estimator
for data in real (unredshifted) space:
the two are related by equation~(\ref{Ws}).
These issues are discussed below.

Suppose that the decision has been made to estimate the power spectrum
$\tilde\xi = \G^\alpha \xi_\alpha$,
equation~(\ref{Gxik}) or equivalently~(\ref{Gxi}),
windowed through some kernel $\G^\alpha$.
How to choose such kernels is the subject of Paper~2,
which illustrates several examples,
such as in Figures~3, 4, and especially Figure 5.
As discussed in \S\ref{minvarpr},
a pair window $W^{ij}$
will give an unbiassed estimate $\hat\xi$,
equation~(\ref{xiest}) or (\ref{Gxiest}),
of the windowed power spectrum $\tilde\xi$
provided that $W^\alpha = \G^\alpha$
(here $W^\alpha$, equation~(\ref{reduce}) or (\ref{reduceD}),
signifies $W^{ij}$ integrated over all pairs $ij$
at given separation $\alpha$ in the survey,
in accordance with the convention established in \S\ref{notation}).
The minimum variance pair window derived in \S\ref{minvarpr} is
$W^{ij} = T^{ij \alpha} T_{\alpha \beta}^{-1} \G^\beta$,
equation~(\ref{minvar}).
But suppose that in place of the true matrix $T$,
it has been decided to use an approximation $\hat T$.
Consider then the approximate pair window defined by
\be
\label{What}
  W^{ij} = \hat T^{ij \alpha} \hat T_{\alpha \beta}^{-1} \G^\beta
  \ .
\ee
Contracting equation~(\ref{What}) with $D^\alpha_{ij}$
(i.e.~integrating over all pairs $ij$ at given separation $\alpha$)
shows that the pair window~(\ref{What}) satisfies
\be
\label{Whata}
  W^\alpha = \G^\alpha
\ee
and therefore yields an unbiassed estimate of $\tilde\xi$,
which is as desired.
The important point here is that the approximation used for
$\hat T^{\alpha \beta}$ in the pair window~(\ref{What})
should be the same as the approximation used for $\hat T^{ij \alpha}$, that is
\be
  \hat T^{\alpha \beta} = D^\alpha_{ij} \hat T^{ij \beta}
  \ .
\ee

The above point may seem rather obvious
-- why would there be any reason not to use the same approximation for both
$\hat T^{ij \alpha}$ and $\hat T^{\alpha \beta}$
in the pair window~(\ref{What})?
The answer lies in the choice of the adjustable constant $k_0$
which enters, via the window $U(\r)$, equation~(\ref{U}),
the approximations proposed in \S\ref{perturbg} for the Gaussian case.
At the end of \S\ref{perturbg} it was suggested that
in evaluating the Fisher matrix $T(k_\alpha,k_\beta)$
it would be appropriate to set
$k_0 = (k_\alpha + k_\beta)/2$,
but in estimating the power spectrum windowed through some kernel
$\G(k)$ narrow about some wavenumber,
it would be appropriate to set $k_0$ equal to that wavenumber.
According to the discussion of the previous paragraph,
for an estimate of the windowed power spectrum to be unbiassed,
it is essential that the same $k_0$ be used for both
$\hat T^{ij \alpha}$ and $\hat T^{\alpha \beta}$
in the pair window~(\ref{What}).
In other words,
one should {\em not}\/ use
$k_0 = (k_\alpha + k_\beta)/2$ for $\hat T^{\alpha \beta}$
in the pair window~(\ref{What}),
even if the kernel $\G(k)$ was constructed in the first place from
a Fisher matrix using that approximation.

It is useful to write out explicitly the relevant equations
for the simplest and most practical case,
where one uses the zeroth order approximation $\hat T = \up{0}{T}$
in the perturbation series of \S\ref{perturbg}, for Gaussian fluctuations.
Suppose that a kernel $\G(k)$ has been chosen
which is suitably narrow in Fourier space about some wavenumber.
For such a kernel,
it would be appropriate to set $k_0$ equal to that wavenumber, a fixed constant.
Section~\ref{perturbg} gives the zeroth order expressions for
$\up{0}{T^{ij \alpha}}$ and $\up{0}{T^{\alpha \beta}}$ in Fourier space.
As long as $k_0$ is taken as a fixed constant, these expressions
can be transformed back into real space, yielding
\be
\label{T0ar}
  \up{0}{T^{ij \alpha}}
    = \frac{1}{2} U(\r_i) U(\r_j) \delta_D(r_{ij} - r_\alpha)
\ee
and
\be
\label{T0abr}
  \up{0}{T^{\alpha \beta}}
    = \frac{1}{2} \delta_D(r_\alpha - r_\beta)
    R(r_\alpha)
\ee
where
\be
\label{R}
  R(r_\alpha)
    =
    \int U(\r_i) U(\r_j) \delta_D(r_{ij} - r_\alpha) 
    \, \ddd r_i \ddd r_j
  \ .
\ee
The quantity $R(r_\alpha)$,
which is often denoted $\langle R R \rangle$ in the literature,
is the expected number of weighted random pairs in the survey
per unit volume interval $4\PI r_\alpha^2 \dd r_\alpha$ of separation
about $r_\alpha$.
It should be emphasized that the expression~(\ref{T0abr}) is {\em not}\/ a
valid general expression for the Fisher matrix;
it is valid only in that the Fourier transform of this expression
is a good approximation to the Fisher matrix $T(k_\alpha,k_\beta)$
for matrix elements satisfying $k_\alpha \approx k_\beta \approx k_0$.
The pair weighting $W^{ij}$ which follows from the zeroth order
approximations~(\ref{T0ar}) and (\ref{T0abr}) is then
\be
\label{W0r}
  W(\r_i,\r_j)
    = {U(\r_i) U(\r_j) \G(r_{ij}) \over R(r_{ij})}
\ee
where $\G(r) = \int \G(k) \e^{-\im\k.\r} \ddd k /(2\PI)^3$
is the kernel $\G(k)$ in the real space representation.
The pair weighting~(\ref{W0r}) has a simple interpretation.
The $U(\r_i) U(\r_j)$ factor, with $U(\r)$ given by equation~(\ref{U}),
is just the FKP pair weighting,
while the $R(r_{ij})$ in the denominator serves as a normalization factor
which ensures the correct overall weighting of pairs at separation $r_{ij}$.
If, as is being assumed, the kernel $\G(k)$ is narrow in Fourier space,
then the kernel $\G(r)$ in real space will presumably be broad and wiggly.
In effect, the pair weighting~(\ref{W0r})
replaces the factor
$\sigma_\alpha^2 j_0(k_\alpha r_{ij}) /2$
in the original FKP pair weighting~(\ref{Wc})
by the factor $\G(r_{ij})/R(r_{ij})$.

The second issue mentioned at the beginning of this section
is that the estimated uncertainty should be the uncertainty in the
actual estimator used, not the uncertainty in the minimum variance estimator.
In general, the expected uncertainty in an estimate $\hat\xi$,
equation~(\ref{xiest}) or (\ref{Gxiest}),
made using an arbitrary pair window $W^{ij}$ is
\be
\label{dxiW}
  \langle \Delta\hat\xi^2 \rangle
    = W^{ij} W^{kl} \langle \Delta\XI_{ij} \Delta\XI_{kl} \rangle
  \ .
\ee
In the particular case of the zeroth order pair window~(\ref{W0r}),
the uncertainty in the corresponding estimator is
\ba
  \langle \Delta\hat\xi^2 \rangle
  \!\!\! &=& \!\!\!
    \int {2 \, \G(r_\alpha)^2 \over R(r_\alpha)} \, \ddd r_\alpha
  \nn
\label{dxi0}
  && \!\!\!
    \mbox{}
    - \int {4 \, \G(r_\alpha) \up{1}{T(r_\alpha,r_\beta)} \G(r_\beta)
    \over R(r_\alpha) R(r_\beta)}
    \, \ddd r_\alpha \ddd r_\beta
\ea
where $\up{1}{T(r_\alpha,r_\beta)}$ on the second line
is the first order correction term~(\ref{T1abg}) to the Fisher matrix,
but expressed in real rather than Fourier space.
Since this perturbation term is small,
a satisfactory approximation to the uncertainty
$\langle \Delta\hat\xi^2 \rangle$ in the estimator should be
\be
\label{dxi00}
  \langle \Delta\hat\xi^2 \rangle
    \approx
    \int {2 \, \G(r_\alpha)^2 \over R(r_\alpha)} \, \ddd r_\alpha
  \ .
\ee
The estimate~(\ref{dxi00}) of the uncertainty increases monotonically
as the constant $\xi(k_0)$ in the pair window increases,
since $U(\r)$, equation~(\ref{U}),
hence $R(r)$, equation~(\ref{R}),
decreases as $\xi(k_0)$ increases.
Thus it might be reasonable
to allow for the small correction to the uncertainty arising from the
perturbation term on the second line of equation~(\ref{dxi0})
by computing the uncertainty from the main term~(\ref{dxi00})
with a value of $\xi(k_0)$ enhanced somewhat over that used in the pair
window $W^{ij}$ itself.
Presumably $\xi(k)$ varies somewhat over the extent of the kernel $\G(k)$,
providing a natural estimate of the amount by which $\xi(k_0)$ should
be enhanced.
This procedure for estimating the uncertainty in the estimator
based on the zeroth order pair window~(\ref{W0r})
is used in Paper~2, Figure~6.

The uncertainty~(\ref{dxi00}) in the approximate estimator should be somewhat
larger, hopefully not by much, than the uncertainty
$T_{\alpha \beta}^{-1} \G^\alpha \G^\beta$
in the minimum variance estimator.
Computing both uncertainties would
give an estimate of how much worse the approximate estimator
is compared to the minimum variance estimator,
and would also provide a useful numerical check.

The third issue remarked at the beginning of this section
is the fact that the minimum variance pair window for data in redshift space
is not the same as the minimum variance pair window for data in real space:
the two are related, at least for linear redshift distortions,
by equation~(\ref{Ws}).
However, equation~(\ref{Ws}) is only part of the solution to the problem
of measuring the power spectrum in optimal fashion from data in redshift space.
A full solution would involve, at least,
simultaneous measurement of the linear distortion parameter
$\ff \approx \Omega^{0.6} / b$.
I hope to examine this problem fully in a subsequent paper.

In the meantime,
it should be remarked that there is no harm in applying an approximate
pair window, such as that given by equation~(\ref{W0r}) for example,
to data in redshift space,
as long as one recognizes that such a pair window
may not necessarily be near minimum variance.
By construction,
the pair window~(\ref{W0r})
applied to densities $\delta^{(s)}_i \delta^{(s)}_j$ in redshift space
will give an unbiassed estimate
of the angle-averaged redshift space power spectrum $\xi^{(s)}(k)$
windowed over the kernel $\G(k)$, for any choice of window $U(\r)$.
However,
the uncertainty in the resulting estimator will not be given by
equation~(\ref{dxiW}),
but rather by
\be
  \langle \Delta\hat\xi^2 \rangle
    = W^{ij} W^{kl} \langle \Delta\XI^{(s)}_{ij} \Delta\XI^{(s)}_{kl} \rangle
\ee
where
$\langle \Delta\XI^{(s)}_{ij} \Delta\XI^{(s)}_{kl} \rangle$
is the covariance of the covariance of the survey in redshift space.
This redshift space covariance
is given in the case of linear redshift distortions by equation~(\ref{XisXis}).

\section{Summary}
\label{summary}

This is the first of two papers
which address the problem of estimating the unredshifted power spectrum
in optimal fashion from a galaxy survey.
Two principal results have been obtained here.

The first result is a derivation of the minimum variance pair weighting,
equation~(\ref{minvar}),
for estimating
the unredshifted power spectrum windowed over some arbitrary prescribed kernel.
The minimum variance pair window is valid for arbitrary survey geometry,
and for Gaussian or non-Gaussian fluctuations.
The generalization of the minimum variance window into redshift space,
for linear redshift distortions, is given in \S\ref{redshift},
although the problem of measuring the linear redshift distortion parameter
$\ff \approx \Omega^{0.6}/b$ itself is not addressed here.

The second principal result is a practical, albeit approximate,
procedure for evaluating the Fisher information matrix $T^{\alpha \beta}$.
The Fisher matrix is the inverse of the expected covariance matrix
of minimum variance estimates of power spectra,
and in effect determines the maximum amount of information about the power
spectrum which can be extracted from a survey.
The proposed procedure for evaluating the Fisher matrix is a perturbation
expansion,
in which the zeroth order approximation is essentially the FKP approximation,
valid for Gaussian fluctuations in the classical limit where the wavelength
is short compared to the scale of the survey.
This zeroth order approximation to the Fisher matrix is used in Paper~2.

\section*{Acknowledgements}

This work was supported by
NSF grant AST93-19977
and by
NASA Astrophysical Theory Grant NAG 5-2797.
I thank Michael Vogeley \& Alex Szalay for sending a preprint of their paper,
which inspired the present paper.
I thank Max Tegmark and the referee, Alan Heavens,
for comments which helped materially to improve the paper.

\end{document}